# Examining the Challenges of Intellectual Property in AI-Generated Productions

**Ali Mazhar***
Attorney at Law, Central Bar Association, Tehran, Iran

**Mohammad Zare**
Artificial Intelligence Laboratory, Arioobarzan Engineering Team, Shiraz, Iran

**Marjan Veysi**
Artificial Intelligence Laboratory, Arioobarzan Engineering Team, Shiraz, Iran

*Corresponding Author: mazhar.law@gmail.com

**Abstract**

With the advancement of artificial intelligence systems capable of autonomously generating artistic, literary, musical works, and even inventions without direct human intervention, the intellectual property (IP) regime faces unprecedented questions and challenges. The most critical issue concerns the ownership of moral and economic rights in the absence of a human creator, and how such outputs can be granted legal protection. This paper first reviews the theoretical foundations and existing literature in this domain, then comparatively examines Iranian legal frameworks—such as the 1969 Law for the Protection of Authors, Composers, and Artists' Rights and the Patent and Trademark Registration Law—alongside other legal systems, including the European Union, the United Kingdom, and the United States. Furthermore, existing legal perspectives on the intellectual property of AI-generated works and the related enforcement challenges are analyzed. The findings reveal significant regulatory gaps within the current Iranian legal framework. To balance the promotion of innovation with the preservation of human creativity, revising existing laws and introducing novel approaches—such as defining a specific intellectual property right for AI-generated works or designating ownership among associated human agents—appears to be essential.

**Keywords:** artificial intelligence, intellectual property, copyright, patent, autonomous works, legal challenges, Iranian law, comparative law

---

## 1. Introduction

The emergence and proliferation of artificial intelligence systems in recent years has led to the production of a new type of intellectual output that is created without direct human intervention. These outputs can include texts generated by language models, images and artworks created with the help of generative networks, musical compositions, or technical inventions designed by artificial intelligence. Such advancements have, on the one hand, created new opportunities for innovation and technological progress; but on the other, they have confronted the legal system with a fundamental question: who is the creator and owner of these works? In the classical intellectual property (IP) system, human intellectual creation forms the basis of legal protection; a work is protectable if it arises from the creativity and thinking of a natural person (Abbott, 2016). If a work is produced by an algorithm or intelligent machine without a human playing a direct role in its creation, can that work still be considered an "intellectual work" and receive legal protection? If so, to whom will its exclusive rights belong—the user who gave the AI a prompt, the company or programmer who created and trained the system, or neither (Gervais, 2019)?

These questions are not merely theoretical; they have also found numerous practical manifestations in recent years. For example, in 2022, a digital painting created entirely by an artificial intelligence (upon receiving a text prompt from a user) won an award at a visual arts competition in the state of Colorado in the United States (Hristov, 2023). The revelation that the direct creator of the work was not a human sparked controversy and again raised the question of how the concept of creator in art and literary law should be interpreted, and to whom the ownership of such a digital work belongs (Hristov, 2023). Further examples can be cited in the domains of text, music, and even invention, where the human role was limited to designing and launching the intelligent system or providing an initial input, while the major part of the creative act was performed by the machine. All of these developments demonstrate the necessity of rethinking traditional IP rules and adapting them to the age of artificial intelligence (Ginsburg & Budiardjo, 2019).

This paper attempts, while reviewing the existing literature and related theoretical foundations, to analyze the Iranian legal framework in confronting this phenomenon and—through a brief comparison with the legal systems of several other countries—to identify the legal deficiencies and challenges. Subsequently, various legal perspectives on the intellectual property of autonomous AI productions are articulated, and practical problems in implementing current laws are discussed. Finally, in the conclusion, solutions and recommendations will be offered for updating regulations so that an appropriate balance can be struck between sup-porting AI-based innovations and preserving the rights and values of human creativity.

## 2. Literature Review and Theoretical Foundations

Intellectual property law is a branch of law concerned with protecting human intellectual and creative works. The primary philosophy of this law is to incentivize creators to produce new works and encourage innovation by granting them temporary exclusive rights. In both major categories of intellectual property—copyright in the domain of literary and artistic works, and patents in the domain of inventions and technical innovations—the fundamental presumption has always been the existence of a human creator or inventor (Gervais, 2019). For example, Article 1 of the Iranian Law for the Protection of Authors, Composers, and Artists' Rights (enacted 1969) provides a definition of "creator" in which, although the word "human" is not explicitly mentioned, the legislative intent is clearly a human creator (Law for the Protection of Authors' Rights, 1969). Similarly, in other legal systems, the principle is that intellectual creation is an act peculiar to natural persons possessing human will and consciousness; a machine or automated process lacks human initiative and feeling and therefore cannot be recognized as a creator or inventor (Abbott, 2016).

However, AI advances have challenged this unwritten traditional principle (Ginsburg & Budiardjo, 2019). From a theoretical standpoint, several different approaches have emerged regarding the possibility or impossibility of protecting AI productions: First, a conservative approach that emphasizes fidelity to classical foundations and asserts that as long as the human element of creativity is absent, no work has come into existence that deserves protection (Hristov, 2023). Based on this view, content produced by AI is merely the product of a technical process and not an "intellectual work" in the legal sense; therefore, no exclusive right over it is conceivable and such outputs fall into the public[1] domain (Gervais, 2019). In contrast, more innovative approaches have appeared that attempt to define roles for humans in the AI production process so that rights can be attached to the final product. Some scholars argue that even if AI has directly created a work, the human user's role in providing the initial idea, configuring the input (prompt), and selecting from among the outputs still constitutes a form of creativity and artistic direction (Yanisky-Ravid & Vardi, 2017). In other words, the human user appears in the role of director or set designer, employing the AI as an advanced tool; he can therefore still be regarded as the creator of the work, even if the instrument of creation has changed (Yanisky-Ravid & Vardi, 2017). This argument draws a fine line between "being a mere tool" and "participating in creation" and places the degree of human creative involvement as the standard for protection (Ginsburg & Budiardjo, 2019).

From the perspective of international bodies, the subject is also under review. The World Intellectual Property Organization (WIPO) in recent years has initiated global conversations about AI and intellectual property and has raised questions, including whether the traditional concepts of inventor and author need to be redefined in light of intelligent technologies, or whether existing interpretations suffice (WIPO, 2020). Reports and expert sessions in this regard indicate that no clear international consensus exists and countries have adopted different approaches (Gheisari et al., 1403 [2024/2025]). Some researchers have even gone further and proposed granting legal personhood to AI so that rights and obligations could be attributed to it (Abbott, 2016); though this idea currently remains theoretical and has many opponents who consider it contrary to the foundations of accountability and ethics. Overall, the theoretical legal foundations in this domain have undergone serious transformation, and a growing body of research is taking shape that this paper attempts to summarize in the following sections.

## 3. Comparative Analysis: Iranian Law and Other Legal Systems

To date, no explicit regulation specifically addressing works produced by artificial intelligence has been enacted in Iranian law (Fattahi, 1402 [2023/2024]); therefore, one must refer to existing intellectual property laws and assess the extent to which they encompass this phenomenon. In the domain of copyright and related rights, the Law for the Protection of Authors, Composers, and Artists' Rights enacted in 1969 is the primary instrument. As mentioned, this law regards the creator as the author, composer, or artist and, while granting them economic and moral rights, had no conception of a work being created by a non-human entity (whether animal or machine) (Law for the Protection of Authors' Rights, 1969). The consequence is that if a work is produced entirely by AI without the creative participation of a human, it is difficult to bring it within the scope of that law's protection. Even concepts such as "joint work" or "derivative work" in this law are all premised on the existence of a human creator (Alipour, 1404 [2025/2026]). On the other hand, the Law on the Protection of the Rights of Computer Software Creators enacted in 1379 [2000/2001], which was specifically enacted to protect software, could to some extent be used to protect the technical aspects of AI products (Law on the Protection of Computer Software Creators' Rights, 1379). This law recognizes the economic and moral rights of the software for its creator and sets the term of economic rights at 30 years from the time the software was created. The argument of some Iranian jurists is that since AI systems are ultimately a type of complex computer software, the designers and developers of such systems can be made beneficiaries of the rights provided under that law (Fattahi, 1402). However, one must note that this argument pertains to the AI system itself (as a piece of software or invention) and not necessarily to the artistic works and content produced by it. In other words, the software law protects the programmer vis-à-vis the code or system designed; but the legal status of the system's independent outputs (such as an image or poem produced by the program) is not clear.

A similar situation is observed in the domain of patent and

[1] public domain

industrial property law. The Patent, Industrial Designs, and Trademarks Registration Law enacted in 1386 [2007/2008] (and its related regulations) defines an invention as the result of the thinking of an individual or individuals who for the first time present a specific product or process. The inventor in legal convention is the very natural person who has performed the mental initiative. The text of this law contains no direct reference to artificial intelligence, and naturally, a non-human inventor was not within the legislator's contemplation (Patent Registration Law, 1386). In 1403 [2024/2025], a new law entitled the Industrial Property Law was enacted to replace the 1386 law, seeking to align Iranian law with international standards (Industrial Property Law, 1403). However, at the time of writing this paper, no specific clause has been reported in the new law that resolves the question of inventor or the intellectual property of AI works; therefore, the principle probably remains that the inventor must be a person with legal capacity (a human or legal entity) and the registration of a patent listing an AI as the inventor will face legal obstacles (Gheisari et al., 1403).

Different approaches are observed in other legal systems. In the United States, the practice of courts and administrative agencies has explicitly excluded works lacking a human creator from copyright coverage (United States Copyright Office, 2022). The US Copyright Office (USCO) has rejected several applications for the registration of fully-automated AI works and has emphasized that a work is protectable only when it is the result of human involvement and creativity (Hristov, 2023). A well-known case was the "monkey selfie" case (Naruto v. Slater), which, although it concerned an animal rather than AI, led the court to rule that a work in which a human played no creative role does not qualify under copyright law. This logic has now been extended to AI works (Ginsburg & Budiardjo, 2019). As a result, much AI-generated content in America is potentially deemed to lack private ownership and falls into the public domain unless a human element of creativity is involved (Yanisky-Ravid & Vardi, 2017).

In contrast, some common law systems like the United Kingdom have sought to fill the existing gap through explicit legislation. The UK Copyright, Designs and Patents Act (1988) provides in one provision that if a work is created by computer without a human author, the person who takes the necessary arrangements for the creation of the work (for example, the programmer or user who provided the necessary input) is recognized as the author of the work (Copyright, Designs and Patents Act 1988, s. 9(3)). This legal innovation means that in England, fully automated works also enjoy at least minimal protection and the rights belong to the human who caused the work to be created (Abbott, 2016). Of course, this rule has not become widespread internationally and most countries do not have a similar provision. The European Union legal system (despite numerous harmonizing regulations on copyright) has not yet adopted a uniform position on AI works and implicitly follows the same traditional principle of a human creator (European Patent Office, 2021). In other words, if a case concerning a fully automated work were brought before European courts, the result would likely be similar to that of the United States and the work would not be found protectable without a human creator (even though no landmark case in this area has yet been brought before the Court of Justice of the European Union) (Gervais, 2019).

In the domain of patents, the story continues in the same vein. In recent years, the famous DABUS case was raised in several countries, in which an inventor and researcher (Dr. Stephen Thaler) attempted to register an invention that had been entirely created by the AI system named DABUS and listed the name of the AI as the inventor in the patent application (Bidar, 1401 [2022/2023]). This request was rejected by the patent offices of the United Kingdom, the European Union, and the United States, and their common reason was that according to the law, an inventor must be a human individual (European Patent Office, 2021). Even Thaler's argument that—as the owner and designer of the AI—patent rights would transfer to him was rejected; because a machine lacks legal personhood and cannot transfer its own rights (Gheisari et al., 1403). In contrast, two exceptional cases occurred: one in South Africa, which for the first time in 2021 issued a patent with an AI listed as the inventor (Bidar, 1401); the other in Australia, where the trial court initially ruled that that country's patent law posed no obstacle to accepting a non-human inventor (Bidar, 1401). Although the Australian ruling was overturned on appeal, and the vast majority of countries now emphasize the necessity of the inventor being human (Gheisari et al., 1403), these exceptional cases have sparked extensive legal debates. In Iranian law, no such case has been reported to date, and ordinarily, if someone were to attempt to register an invention in the name of an AI, the Industrial Property Office would not accept it. Under the general principles of the Iranian Civil Code, legal capacity for rights is recognized exclusively for persons (natural and legal), and property and rights cannot be ownerless or belonging to an object lacking personhood; therefore, AI—which is neither a human nor a legal entity—cannot itself be regarded as an owner or right-holder (Iranian Civil Code). This general principle, combined with the absence of explicit statutory provision, requires that patent or copyright rights over AI outputs either be allocated to a person with legal capacity (such as a related human inventor/author) or simply not come into existence, with the work falling into the public domain.

## 4. Legal Analysis and Existing Perspectives

Given the state of existing laws, jurists and policymakers have identified three main approaches in confronting the issue of intellectual property in AI productions (Yanisky-Ravid & Vardi, 2017):

1. **Developer or owner of the AI system as owner of the work:** According to this view, the company that designed and trained the AI model—or in general, the person who played the primary financial and technical role in creating the system—should be considered the owner of its outputs (Gervais, 2019). The justification for this approach is that AI is a tool in the hands of its maker and the final product is in fact the result of its investment and initiative. Especially in cases where the system's output is directly the product of its trained architecture and data, the user's creative role appears minimal. However, critics of this approach argue that such a view ignores the creativity of the user or client (Hristov, 2023). A user who creatively employs an AI to produce an original work may have played no lesser role than the developer in the emergence of the work. Furthermore, giving all rights to the companies owning the technology could lead to a concentration of power in the hands of a small number of large companies and reduce the incentive of individual users.

2. **User (prompt designer) as creator or owner of the work:** This view, which has gained many adherents, considers the human user who provides the commands, idea, or initial data to the system to be deserving of ownership of the work (Ginsburg & Budiardjo, 2019). The argument of supporters of this theory, as noted in the literature review, is that working with an advanced AI is itself an art and a form of creativity; the precise selection of words or parameters, repeatedly refining the input to reach the desired result, and ultimately selecting the best output from among the results obtained all require initiative, taste, and human effort (Yanisky-Ravid & Vardi, 2017). Thus the final work, although produced through the machine, is the manifestation of indirect human creativity and its rights should belong to that person. The fundamental challenge in this approach is determining how much involvement and initiative from the user is required for that person to be genuinely regarded as the creator of the work (Hristov, 2023). Is merely pressing a button and randomly generating an image sufficient, or must the user play an active role in guiding the process? This requires the formulation of clear legal (and technical) criteria. Some have suggested that if the AI output is predictable and under the complete control of the user, then the user may be considered the creator; but if the system autonomously and unpredictably produces a work (so that the user is merely observing the result), the user cannot be deemed the creator. Drawing this line is difficult in practice and carries the potential for litigation.
3. **Non-recognition of a specific owner (public domain):** The third view holds that when the direct creator of a work is not human, no exclusive right arises at all and the work should, like an ownerless work, be placed in the public domain (Abbott, 2016). Proponents of this theory invoke the traditional foundations of copyright, which viewed exclusive protection as an instrument for incentivizing human creativity. In the case of AI, it is argued that AI itself needs no incentive, and the developers and users of AI also have other motivations (including commercial or scientific interests) for using AI. Therefore, if AI outputs enter the public domain, human knowledge and culture will be enriched without any harm to innovation. The primary criticism of this approach is that in practice it may reduce the incentive for investment in AI projects (Gervais, 2019). Companies and individuals invest heavily in creative AI systems when they hope to derive benefits from the results. If no protection whatsoever is envisaged over the results and others can immediately copy and exploit them, the pace of innovation may slow. Furthermore, this extreme view may not be considered fair either, because it leaves the indirect role of human creators and users completely unrewarded.

In addition to discussions about the ownership of the work, other legal aspects of AI works are also noteworthy. One of these is the question of the moral rights[2] of the creator, recognized in Iranian law and many other countries (such as the right to attribution of the work to the creator's name and the right to integrity). Granting these rights to an AI is meaningless, as it lacks human personhood and the sense of literary-artistic honor and reputation. On the other hand, if we wish to designate the user or developer as the presumptive author, is it appropriate, for example, to publish a work that was practically created by AI under someone else's name? This has been questioned from the perspective of scholarly and artistic integrity. In practice, many users present AI outputs under their own names; but in the legal framework, attributing a work to a person who was not its actual creator raises ethical and legal doubts. Some jurists have proposed that if AI works are to receive protection, only economic rights should be granted and traditional moral rights eliminated or applied in a more limited form (Gheisari et al., 1403; Ginsburg & Budiardjo, 2019).

Another important matter is legal liability in respect of potential infringement of others' rights by AI. Consider a scenario in which an AI system, in its learning process, uses copyrighted works without permission and then produces an output similar to the original. Such a matter has already become the subject of important legal actions; for example, Getty Images and a group of artists have filed a complaint against the developers of image-generating models (such as Stable Diffusion), alleging that they used their copyrighted works to train the algorithm and then generate images similar to those styles (Hristov, 2023). These actions have pushed the debate beyond the ownership of outputs to the legitimacy of input data and AI learning methods. Although that subject itself merits separate extended treatment, its relevance to the present discussion lies in the fact that AI outputs may inadvertently infringe the rights of other rights-holders. In such a case, who is held responsible—the user who issued the production command, the company that built the algorithm, or neither (on the argument that the act proceeded directly from the machine)? Answers to these questions are still at an early stage of discussion, but legal systems will likely follow the same general approach they apply in other cases of product and tool liability: indirect or fault-based liability will fall on the humans who had control of or operated the AI (such as the producer or user of a defective product). In any case, the enforcement challenges related to identifying instances of infringement and determining responsibility in

[2] moral rights

the domain of AI works are far more complex than in traditional cases and will require the development of clear rules in the future (Yanisky-Ravid & Vardi, 2017).

## 5. Enforcement and Legal Challenges

The analyses above show that the legal vacuum and conceptual ambiguity in the intellectual property of AI works can lead to numerous practical challenges. In this section, some of the most important of these challenges are identified across two dimensions—enforcement and legal:

- **Legal uncertainty and increased disputes:** The absence of clear laws can itself increase the incentive for litigation. For example, if two different persons claim ownership of a work produced by an AI (one as the algorithm's creator and another as the user-creator of the work), courts must decide on the basis of analogies from existing laws or general legal doctrines, which may reach different outcomes in similar cases. Such a situation causes confusion among stakeholders and reduces legal predictability, which is itself an obstacle to innovation (Gervais, 2019). Furthermore, if clear judicial precedent does not emerge, some individuals or companies may exploit the legal vacuum and infringe on others' rights (for example, by freely copying and exploiting AI-produced works of others on the pretext that they are unprotected). Consequently, any delay in reforming regulations can have adverse practical consequences.
- **Administrative problems in registering and protecting rights:** Current registration systems (such as patent offices or book and software registration systems at the Ministry of Culture and Islamic Guidance) are designed around the identity information of humans. If someone today applies to register an AI-produced artwork and writes "an AI program" in the creator field of the application form, the request will likely be rejected (Bidar, 1401). Similarly, in patent registration, stating the name of a human inventor is mandatory (Gheisari et al., 1403). This causes some individuals to be compelled to submit false information (for example, stating their own name or that of a colleague instead of the AI as the inventor) in order for their invention to be registrable. Although such workarounds nominally confer rights, they are not desirable from the perspective of ethical principles and legal transparency, and could in the future call into question the validity of registered rights (for example, if it is revealed that the actual inventor was not human). Therefore, administrative mechanisms need to be adapted and updated to take account of the particular circumstances of AI works—whether through new forms or explicit implementing guidelines.
- **Evaluating criteria of originality and creativity:** In copyright law, originality[3] is the condition for protecting a work. Assessing originality ordinarily means that the work arises from the independent effort and creativity of the creator and is not copied from another. For AI works, a complex question is: to whom should originality be attributed? If the human user determined the general idea or style and the AI processed the details, is the resulting work original? On the one hand yes, because the output is unique and novel; but on the other hand, because it is not directly born of the human mind, it may not be considered original (Ginsburg & Budiardjo, 2019). Furthermore, as AI advances, distinguishing AI-produced works from human-made ones becomes more difficult. This will be challenging for examiners at registration offices and even for courts in drawing the line between a protectable original work and a mere mechanical combination of data.
- **Volume of outputs and potential for abuse:** One of the characteristics of generative AI is that it can produce a massive volume of outputs in a short time. Imagine one person using AI to create thousands of different images or texts in a single day. If these are protectable, can they claim copyright over all of them? This could impose a heavy burden on the system for registering and enforcing intellectual property rights and distort the concept of a work's uniqueness—which has traditionally been linked to the effort and toil of its creator. There are also concerns from artists and writers about an uneven competitive landscape: that is, individuals with the help of AI could produce a flood of works without spending much creative time and claim the rights to them, while human creators must invest time and genius in each work. This is also a concern from the perspective of cultural policy and the protection of human creativity (Hristov, 2023).
- **Cross-border dimension and conflict of laws:** Finally, it must be noted that the question of AI works is a global phenomenon and the outputs of a system can spread rapidly around the world. Divergence of country approaches (for example, if one country deems a work unprotected and another considers it covered by copyright) can lead to confusion in the observance of rights between countries. For example, a work produced in Iran and unprotected here might be registered and claimed in Europe by someone else, or vice versa. Such a situation recalls the necessity of international coordination and continuous dialogue among legislators so that common standards can be found—at least as far as possible—for determining the status of such works (WIPO, 2020).

## 6. Conclusion and Recommendations

Intellectual property in autonomous AI productions is one of the most challenging legal subjects of the present age, with broad legal, economic, and philosophical dimensions (Abbott, 2016). On the one hand, ignoring and refusing to protect such works can lead to chaos and a decrease in innovators' incentives; as individuals and companies, in the absence of legal protection, may have no inclination

[3] originality

to publish or invest in creative AI-based projects (Gervais, 2019). On the other hand, granting full protection equal to entirely human works is also not without risk; there is a concern that the value and standing of human creativity may be diminished and that a flood of machine-produced works may confront the intellectual property system with an unmanageable volume (Hristov, 2023). The desirable solution, therefore, is to adopt an approach that neither hinders technological progress nor abandons its achievements, but rather regulates rights and obligations in a forward-looking and orderly manner (Ginsburg & Budiardjo, 2019).

For the Iranian legal system—whose current laws do not explicitly address this problem (Fattahi, 1402)—the following measures are recommended:

1. **Revision of copyright law and related statutes:** It is necessary for the legislature, by forming specialized working groups, to update the 1969 Law and other regulations and to clarify the status of works without a human creator. One option is to draw inspiration from the British experience, by specifying in the law that in the absence of a human creator, the person who played the decisive role in creating the work (whether programmer, user, or primary investor) shall be deemed the creator. Admittedly, identifying that person in each case may be difficult and may require implementing regulations or supplementary guidelines (Gheisari et al., 1403).
2. **Creating a special right or quasi-copyright for AI works:** As some experts have proposed (Yanisky-Ravid & Vardi, 2017), a new category within the IP system can be defined that is specific to works produced by intelligent systems. This right may be more limited than ordinary copyright in both term of protection and scope of rights, so that both the necessary incentive for innovation remains and short-term monopoly does not prevent long-term public exploitation. For example, a non-renewable protection of 5 or 10 years may be sufficient for these works. Additionally, some moral rights such as the right of attribution can be modified (for example, requiring the work to be identified as AI-generated alongside the name of the rights-holder).
3. **Developing practice in the domain of AI patents:** It is recommended that the Iranian Patent Office, monitoring global developments, issue a guideline specifying how the inventor should be declared if an invention was created with the involvement of AI. Perhaps the best approach is still to insist on listing the name of one or several human inventors (as persons who conceived the initial idea or set up the project) so that the registered patent is not legally vulnerable. At the same time, it is necessary to add a section to patent registration forms explaining the degree to which AI was used in the invention. This transparency helps ensure that if the law changes or a dispute arises in the future, a clear record of AI's role in the patent file is available (Bidar, 1401).
4. **Training and capacity building:** The judicial and administrative system needs to upgrade its technical and legal knowledge to confront such novel cases. Training judges, court experts, and IP office examiners on the fundamentals of AI technology and how it functions is essential so they can better understand the distinction between the human and machine contributions in creating works. Furthermore, inviting technology experts to participate in drafting new laws and regulations will lead to more operationally viable solutions (WIPO, 2020).
5. **International cooperation:** The issue of AI knows no borders, and it is therefore necessary for Iran to participate actively in international forums (such as WIPO) to play a role in the process by which global rules and norms are shaped. Exchange of experience with other countries—especially countries with similar legal systems (for example, countries with civil law traditions inspired by French law)—can be beneficial. It is to be hoped that in the future, an international treaty or at minimum recommendations on the rights of AI works will be developed, and it is preferable that Iran be among the pioneers so that the interests of domestic stakeholders are also taken into account (WIPO, 2020).

In closing, it must be emphasized that the AI age brings opportunities as well as challenges for the legal system. Just as previous technologies (printing, photography, the computer, and the internet) brought about transformations in legal thinking and laws ultimately adapted to them, AI has similarly prompted law to strive for innovation and dynamism. Appropriate regulation in this domain can, while supporting the growth of technology and new creativity, perpetuate at a new level the fundamental values of intellectual property law—honoring human creativity and encouraging innovation (Abbott, 2016).

---

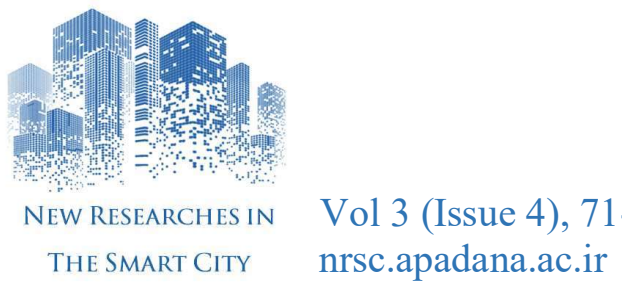

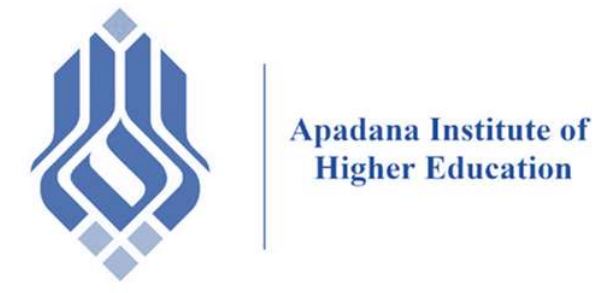

# Examining the Challenges of Intellectual Property in AI-Generated Productions

**Ali Mazhar** *

Attorney at Law, Central Bar Association, Tehran, Iran.

**Mohammad Zare**

Artificial Intelligence Laboratory, Arioobarzan Engineering Team, Shiraz, Iran.

**Marjan Veysi**

Artificial Intelligence Laboratory, Arioobarzan Engineering Team, Shiraz, Iran.

## Abstract

With the advancement of artificial intelligence systems capable of autonomously generating artistic, literary, musical works, and even inventions without direct human intervention, the intellectual property (IP) regime faces unprecedented questions and challenges. The most critical issue concerns the ownership of moral and economic rights in the absence of a human creator, and how such outputs can be granted legal protection. This paper first reviews the theoretical foundations and existing literature in this domain, then comparatively examines Iranian legal frameworks—such as the 1969 Law for the Protection of Authors, Composers, and Artists' Rights and the Patent and Trademark Registration Law—alongside other legal systems, including the European Union, the United Kingdom, and the United States. Furthermore, existing legal perspectives on the intellectual property of AI-generated works and the related enforcement challenges are analyzed. The findings reveal significant regulatory gaps within the current Iranian legal framework. To balance the promotion of innovation with the preservation of human creativity, revising existing laws and introducing novel approaches—such as defining a specific intellectual property right for AI-generated works or designating ownership among associated human agents—appears to be essential.

* Corresponding Author: mazhar.law@gmail.com

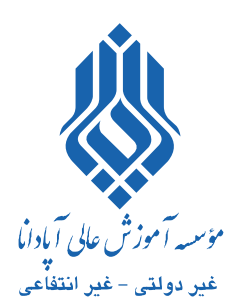

# بررسی چالش‌های مالکیت فکری در تولیدات خودکار توسط هوش مصنوعی

**علی مظهر** * وکیل پایه یک دادگستری، کانون وکلای دادگستری مرکز، تهران، ایران.

**محمد زارع** آزمایشگاه هوش مصنوعی، تیم مهندسی آریوبرزن، شیراز، ایران.

**مرجان ویسی** آزمایشگاه هوش مصنوعی، تیم مهندسی آریوبرزن، شیراز، ایران.

## چکیده

با پیشرفت سیستم‌های هوش مصنوعی که قادر به تولید خودکار آثار هنری، ادبی، موسیقیایی و حتی اختراعات بدون دخالت مستقیم انسان هستند، نظام حقوق مالکیت فکری با پرسش‌ها و چالش‌های بی‌سابقه‌ای روبه‌رو شده است. مهم‌ترین مسئله این است که در غیاب خالق انسانی، حقوق مادی و معنوی این تولیدات به چه کسی تعلق می‌گیرد و چگونه می‌توان از آن‌ها حمایت حقوقی کرد؟ این مقاله، ابتدا به مبانی نظری و مرور ادبیات در این حوزه می‌پردازد و سپس، مقررات حقوقی ایران (ازجمله قانون حمایت حقوق مؤلفان مصوب ۱۳۴۸ و قانون ثبت اختراعات و علائم تجاری) را در کنار نظام‌های حقوقی دیگر (مانند اتحادیه اروپا، بریتانیا و ایالات‌متحده) به‌صورت تطبیقی بررسی می‌کند. در ادامه، دیدگاه‌های حقوقی موجود در خصوص مالکیت فکری آثار تولیدشده توسط هوش مصنوعی و چالش‌های اجرایی مرتبط تحلیل می‌گردد. نتایج حاکی از آن است که خلأهای مقرراتی قابل‌توجهی در حقوق کنونی ایران وجود دارد به‌نحوی‌که برای حفظ تعادل بین تشویق نوآوری و حفظ ارزش خلاقیت انسانی، بازنگری قوانین و ارائه راهکارهای نوین مانند تعریف یک حق مالکیت فکری ویژه برای آثار هوش مصنوعی یا تعیین مالک از میان عوامل انسانی مرتبط، ضروری به نظر می‌رسد.

* نویسندهٔ مسئول: mazhar.law@gmail.com

## مقدمه

ظهور و گسترش سامانه‌های هوش مصنوعی در سال‌های اخیر موجب تولید نوع جدیدی از خروجی‌های فکری شده است که بدون مداخله مستقیم انسان خلق می‌شوند. این خروجی‌ها می‌توانند شامل متن‌هایی که توسط مدل‌های زبانی تولید می‌شوند، تصاویر و آثار هنری خلق‌شده به کمک شبکه‌های مولد، قطعات موسیقی یا اختراعات فنی طراحی‌شده توسط هوش مصنوعی باشند. چنین پیشرفت‌هایی از یک سو فرصت‌های تازه‌ای برای نوآوری و پیشرفت تکنولوژیک فراهم کرده است؛ اما از سوی دیگر نظام حقوقی را با پرسشی بنیادین مواجه ساخته است: چه کسی خالق و مالک این آثار است؟ در نظام کلاسیک حقوق مالکیت فکری، آفرینش فکری انسان اساس حمایت حقوقی است؛ اثری قابل حمایت است که ناشی از خلاقیت و تفکر یک شخص حقیقی باشد (Abbott, 2016). حال اگر اثری توسط یک الگوریتم یا ماشین هوشمند تولید شود و انسان نقش مستقیمی در خلق آن نداشته باشد، آیا آن اثر کماکان «اثر فکری» محسوب می‌گردد و تحت حمایت حقوقی قرار می‌گیرد؟ اگر آری، حقوق انحصاری آن متعلق به چه کسی خواهد بود؟ کاربری که دستور (پرامپت) را به هوش مصنوعی داده، شرکت یا برنامه‌نویسی که سیستم را ایجاد و آموزش داده یا هیچ‌کدام (Gervais, 2019)؟

این پرسش‌ها صرفاً نظری نیستند بلکه در سال‌های اخیر مصادیق عملی متعددی نیز پیدا کرده‌اند. برای نمونه، در سال ۲۰۲۲ یک نقاشی دیجیتال که تماماً توسط هوش مصنوعی (با دریافت دستور متنی از سوی یک کاربر) خلق شده بود، برنده مسابقه هنرهای تجسمی در ایالت کلرادو آمریکا شد (Hristov, 2023). افشای این موضوع که خالق مستقیم اثر یک انسان نبوده است جنجال‌هایی را در پی داشت و بار دیگر این مسئله را مطرح کرد که مفهوم خالق در حقوق هنر و ادبیات چگونه باید تفسیر شود و مالکیت چنین اثر دیجیتالی متعلق به کیست (Hristov, 2023). نمونه‌های دیگری در حوزه متن و موسیقی و حتی اختراع نیز قابل اشاره‌اند که در آن‌ها نقش انسان صرفاً در طراحی و راه‌اندازی سیستم هوشمند یا ارائه ورودی اولیه خلاصه شده و بخش اعظم خلق اثر توسط ماشین انجام گرفته است. تمامی این تحولات، ضرورت بازاندیشی در قواعد حقوق مالکیت فکری سنتی و تطبیق آن‌ها با عصر هوش مصنوعی را نشان می‌دهد (Ginsburg & Budiardjo, 2019).

در این مقاله، تلاش می‌شود ضمن بررسی ادبیات موجود و مبانی نظری مرتبط، چارچوب حقوقی ایران در مواجهه با این پدیده تحلیل شود و با مقایسه‌ای کوتاه با نظام حقوقی برخی کشورهای دیگر، کاستی‌ها و چالش‌های قانونی مشخص گردد. سپس دیدگاه‌های حقوقی مختلف درباره مالکیت فکری تولیدات خودکار هوش مصنوعی تبیین شده و مشکلات عملی در اجرای قوانین کنونی بحث می‌شود. نهایتاً در بخش نتیجه‌گیری، راهکارها و پیشنهادهایی برای به‌روزرسانی مقررات ارائه خواهد شد تا بتوان توازنی مناسب میان حمایت از نوآوری‌های مبتنی بر هوش مصنوعی و حفظ حقوق و ارزش‌های خلاقیت انسانی برقرار نمود.

## مرور ادبیات و مبانی نظری

حقوق مالکیت فکری شاخه‌ای از حقوق است که به حمایت از آثار فکری و خلاقانه انسان می‌پردازد. فلسفه اصلی این حقوق آن است که با اعطای حقوق انحصاری موقت به پدیدآورندگان، آن‌ها را به خلق آثار جدید ترغیب کند و نوآوری را تشویق نماید. در دو دسته مهم مالکیت فکری یعنی حق مؤلف (کپی‌رایت) در حوزه آثار ادبی و هنری و حق اختراع در حوزه اختراعات و ابتکارات فنی همواره فرض بنیادین بر وجود یک خالق یا مخترع انسان بوده است (Gervais, 2019). برای مثال، ماده ۱ قانون حمایت حقوق مؤلفان و مصنفان و هنرمندان ایران (مصوب ۱۳۴۸) تعریفی از «پدیدآورنده» ارائه می‌دهد که هرچند صراحتاً کلمه انسان در آن ذکر نشده؛ اما به‌وضوح مراد قانون‌گذار خالق

بشری بوده است (قانون حمایت حقوق مؤلفان، ۱۳۴۸). همچنین در نظام‌های حقوقی دیگر نیز اصل بر این است که آفرینش فکری عملی است مختص اشخاص حقیقی دارای اراده و شعور انسانی؛ یک ماشین یا فرایند خودکار فاقد ابتکار و احساس بشری است و لذا نمی‌تواند به‌عنوان پدیدآورنده یا مخترع شناسایی شود (Abbott, 2016).

بااین‌حال، پیشرفت‌های هوش مصنوعی این اصل سنتی نانوشته را به چالش کشیده‌اند (Ginsburg & Budiardjo, 2019). از دیدگاه نظری، چند رویکرد مختلف نسبت به امکان یا عدم امکان حمایت از تولیدات هوش مصنوعی شکل گرفته است: نخست، رویکردی محافظه‌کارانه که بر وفاداری به مبانی کلاسیک تأکید دارد و بیان می‌دارد تا زمانی که عنصر انسانیِ خلاقیت در کار نباشد، اثری به وجود نیامده است که شایسته حمایت باشد (Hristov, 2023). بر مبنای این دیدگاه، محتوای تولیدشده توسط هوش مصنوعی صرفاً محصول یک فرایند تکنیکی است و نه «اثر فکری» به معنای حقوقی؛ لذا هیچ حق انحصاری نسبت به آن قابل تصور نیست و چنین خروجی‌هایی در حوزه عمومی[1] قرار می‌گیرند (Gervais, 2019). در مقابل، رویکردهای نوآورانه‌تری ظاهر شده که تلاش دارند نقش‌هایی برای انسان در فرآیند تولید اثر توسط هوش مصنوعی تعریف کنند تا از این رهگذر بتوان حقوقی را به محصول نهایی تعلق داد. برخی صاحب‌نظران استدلال می‌کنند که حتی اگر هوش مصنوعی مستقیماً دست به خلق اثر زده، باز هم نقش کاربر انسانی در ارائه ایده اولیه یا تنظیم ورودی (پرامپت) و انتخاب از میان خروجی‌ها نوعی خلاقیت و هدایت هنری محسوب می‌شود (Yanisky-Ravid & Vardi, 2017). به بیان دیگر، انسان کاربر در نقش کارگردان یا طراح صحنه ظاهر شده و هوش مصنوعی را مانند یک ابزار پیشرفته به کار گرفته است؛ بنابراین می‌توان او را همچنان خالق اثر دانست، هرچند ابزار خلق تغییر کرده است (Yanisky-Ravid & Vardi, 2017). این استدلال مرز باریکی میان «ابزار بودن صرف» و «مشارکت در خلق» ترسیم می‌کند و میزان دخالت خلاقانه انسان را معیار حمایت قرار می‌دهد (Ginsburg & Budiardjo, 2019).

از منظر نهادهای بین‌المللی نیز موضوع در حال بررسی است. سازمان جهانی مالکیت فکری (WIPO) در سال‌های اخیر گفت‌وگوهای جهانی پیرامون هوش مصنوعی و مالکیت فکری را آغاز کرده و پرسش‌هایی را مطرح نموده است، ازجمله اینکه آیا مفاهیم سنتی مخترع و مؤلف نیاز به بازتعریف در پرتو فناوری‌های هوشمند دارند یا می‌توان با تفسیر موجود نیز پاسخگو بود (WIPO, 2020). گزارش‌ها و نشست‌های تخصصی در این زمینه حاکی از آن است که در سطح جهانی اجماع روشنی وجود ندارد و کشورها رویکردهای متفاوتی اتخاذ کرده‌اند (قیصری و همکاران، ۱۴۰۳). برخی پژوهشگران حتی پا را فراتر گذاشته و شخصیت حقوقی بخشیدن به هوش مصنوعی را مطرح کرده‌اند تا از این طریق بتوان حقوق و تکالیفی برای آن متصور شد (Abbott, 2016)؛ هرچند این ایده فعلاً جنبه نظری داشته و مخالفان زیادی نیز دارد که آن را مغایر بنیان‌های مسئولیت‌پذیری و اخلاق می‌دانند. درمجموع، مبانی نظری حقوقی در این حوزه دچار تحولی جدی شده و ادبیات پژوهشی رو به رشدی در حال شکل‌گیری است که این مقاله سعی دارد چکیده‌ای از آن را در ادامه ارائه نماید.

## بررسی تطبیقی: حقوق ایران و نظام‌های دیگر

در حقوق ایران تاکنون مقرره صریحی که به‌طور خاص به آثار تولیدشده توسط هوش مصنوعی بپردازد وضع نشده است (فتاحی، ۱۴۰۲)؛ بنابراین، باید به قوانین موجود در حوزه مالکیت فکری رجوع کرد و میزان شمول آن‌ها را نسبت به این پدیده سنجید. در زمینه حقوق مؤلف و حقوق مجاور، قانون حمایت از حقوق مؤلفان، مصنفان و هنرمندان مصوب ۱۳۴۸ محور اصلی است. همان‌گونه که اشاره شد، این قانون پدیدآورنده را مؤلف، مصنف یا هنرمند دانسته و

1. public domain

ضمن اعطای حقوق مادی و معنوی به وی، هیچ تصوری از خلق اثر توسط غیر انسان (اعم از حیوان یا ماشین) نداشته است (قانون حمایت حقوق مؤلفان، ۱۳۴۸). نتیجه آنکه اگر اثری کاملاً توسط هوش مصنوعی و بدون مشارکت خلاقانه انسان تولید شود، به‌دشواری می‌توان آن را مشمول حمایت این قانون دانست. حتی مفاهیمی نظیر «اثر مشترک» یا «اثر اقتباسی» نیز در این قانون همگی بر فرض وجود خالق انسانی بنا شده‌اند (علیپور، ۱۴۰۲). از سوی دیگر، قانون حمایت از حقوق پدیدآورندگان نرم‌افزارهای رایانه‌ای مصوب ۱۳۷۹ که به‌طور خاص برای حمایت از نرم‌افزار وضع شده تا حدودی می‌تواند برای حمایت از جنبه‌های فنی محصولات هوش مصنوعی به کار رود (قانون حمایت از حقوق پدیدآورندگان نرم‌افزارهای رایانه‌ای، ۱۳۷۹). این قانون حقوق مادی و معنوی نرم‌افزار را برای پدیدآورنده آن به رسمیت شناخته و مدت حقوق مادی را ۳۰ سال از زمان ایجاد نرم‌افزار تعیین کرده است. استناد برخی حقوق‌دانان ایرانی بر این است که چون سیستم‌های هوش مصنوعی درنهایت یک نوع نرم‌افزار رایانه‌ای پیچیده هستند، پس می‌توان طراحان و توسعه‌دهندگان آن‌ها را تحت شمول قانون مذکور از حقوق خود بهره‌مند ساخت (فتاحی، ۱۴۰۲). البته باید توجه داشت که این استدلال ناظر به خود سیستم هوش مصنوعی است (به‌عنوان یک نرم‌افزار یا اختراع) و نه لزوماً آثار هنری و محتوای تولیدشده توسط آن. به بیان دیگر، قانون نرم‌افزار از برنامه‌نویس در قبال کد یا سیستم طراحی‌شده حمایت می‌کند؛ اما تکلیف حقوقی خروجی‌های مستقل سیستم (مثلاً یک تصویر یا شعر ساخته‌شده توسط برنامه) روشن نیست.

در حوزه حقوق اختراعات و مالکیت صنعتی نیز وضعیت مشابهی مشاهده می‌شود. قانون ثبت اختراعات، طرح‌های صنعتی و علائم تجاری مصوب ۱۳۸۶ (و آیین‌نامه‌های مرتبط با آن) اختراع را حاصل فکر فرد یا افراد دانسته که برای اولین بار فرآورده یا فرآیندی خاص را ارائه می‌کند. مخترع در عرف حقوقی همان شخص حقیقی‌ای است که ابتکار ذهنی را انجام داده است. در متن این قانون اشاره مستقیمی به هوش مصنوعی وجود ندارد و طبیعتاً مخترع غیر انسان در منظر قانون‌گذار نبوده است (قانون ثبت اختراعات، ۱۳۸۶). در سال ۱۴۰۳ قانون جدیدی تحت عنوان قانون مالکیت صنعتی به تصویب رسیده که جایگزین قانون ۱۳۸۶ گردیده و تلاش شده تا قوانین ایران را با استانداردهای بین‌المللی هماهنگ‌تر کند (قانون مالکیت صنعتی، ۱۴۰۳). بااین‌حال، تا زمان نگارش این مقاله هیچ بند مشخصی در قانون جدید نیز گزارش نشده که موضوع مخترع یا مالکیت فکری آثارِ هوش مصنوعی را حل کرده باشد؛ بنابراین، احتمالاً همچنان اصل بر آن است که مخترع باید شخص واجد اهلیت (انسان یا شخص حقوقی) باشد و ثبت اختراعی که مخترع آن یک هوش مصنوعی ذکر شده باشد، با موانع قانونی مواجه خواهد شد (قیصری و همکاران، ۱۴۰۳).

در نظام‌های حقوقی دیگر رویکردهای گوناگونی مشاهده می‌شود. در ایالات‌متحده آمریکا، رویه ادارات و محاکم صراحتاً آثار فاقد خالق انسانی را از شمول کپی‌رایت خارج دانسته‌اند (United States Copyright Office, 2022). اداره کپی‌رایت آمریکا (USCO) در چندین مورد درخواست ثبت آثار تمام‌خودکارِ هوش مصنوعی را رد کرده و تأکید نموده که اثر زمانی قابل حفاظت است که حاصل دخالت و خلاقیت انسانی باشد (Hristov, 2023). نمونه معروف، پرونده عکس سلفی میمون (Naruto v. Slater) بود که هرچند مربوط به حیوان بود نه هوش مصنوعی، اما دادگاه حکم کرد اثری که انسان در آن نقش خالق نداشته باشد مشمول قانون کپی‌رایت نیست. این منطق اکنون به آثار هوش مصنوعی نیز تسری داده شده است (Ginsburg & Budiardjo, 2019). درنتیجه، بسیاری از محتوای تولیدشده توسط هوش مصنوعی در آمریکا بالقوه فاقد مالکیت خصوصی تلقی شده و در حوزه عمومی قرار می‌گیرد مگر آن‌که عنصر خلاقیت بشری در آن دخیل گردد (Yanisky-Ravid & Vardi, 2017). در مقابل، در برخی نظام‌های کامن‌لا مانند بریتانیا تلاش شده از طریق قانون‌گذاری صریح، خلأ موجود پر شود. قانون کپی‌رایت بریتانیا (قانون حق مؤلف، طرح‌ها و اختراعات ۱۹۸۸) در ماده‌ای مقرر داشته که اگر اثری به‌صورت رایانه‌ای و بدون

خالق انسانی ایجاد شود، شخصی که اقدامات ضروری برای تولید اثر را انجام داده (برای مثال برنامه‌نویس یا کاربری که ورودی لازم را فراهم کرده) به‌عنوان پدیدآورنده اثر شناخته می‌شود (Copyright, Designs and Patents Act 1988, s. 9(3)). این ابتکاری قانونی بدین معناست که در انگلستان آثار تمام‌خودکار نیز از حداقلی از حمایت برخوردارند و حقوق آن به انسانی که منجر به ایجاد اثر شده تعلق می‌گیرد (Abbott, 2016). البته، این قاعده در سطح بین‌المللی چندان رایج نشده و بیشتر کشورها قاعده‌ای مشابه ندارند. نظام حقوقی اتحادیه اروپا (با وجود مقررات هماهنگ‌کننده متعدد در زمینه کپی‌رایت) تاکنون موضع واحدی در قبال آثار هوش مصنوعی اتخاذ نکرده است و به‌طور ضمنی همان اصل سنتی خالق انسانی را دنبال می‌کند (European Patent Office, 2021). به عبارت دیگر، اگر پرونده‌ای مربوط به اثرِ تمام‌خودکار به دادگاه‌های اروپا برسد، احتمالاً نتیجه مشابه آمریکا خواهد داشت و اثر بدون خالق انسانی، قابل حمایت تشخیص داده نخواهد شد (اگرچه هنوز پرونده شاخصی در این زمینه در سطح دادگستری اتحادیه اروپا مطرح نشده است) (Gervais, 2019).

در زمینه اختراعات، داستان به همین صورت دنبال می‌شود. در سال‌های اخیر پرونده مشهوری به نام DABUS در کشورهای متعددی مطرح شد که طی آن یک مخترع و پژوهشگر (دکتر استفان تالر) تلاش کرد اختراعی را که کاملاً توسط سیستم هوش مصنوعی به نام DABUS خلق شده بود ثبت کند و نام هوش مصنوعی را به‌عنوان مخترع در اظهارنامه ثبت اختراع ذکر نماید (بیدار، ۱۴۰۱). این درخواست در ادارات اختراع انگلستان، اتحادیه اروپا و آمریکا رد شد و دلیل مشترک آن‌ها این بود که طبق قانون، مخترع باید یک فرد انسانی باشد (European Patent Office, 2021). حتی استدلال تالر مبنی بر اینکه به‌عنوان مالک و طراح هوش مصنوعی، حقوق اختراع به او منتقل می‌شود پذیرفته نشد؛ چراکه ماشین فاقد شخصیت حقوقی است و نمی‌تواند حقوق خود را انتقال دهد (قیصری و همکاران، ۱۴۰۳). در مقابل، دو مورد استثنائی رخ داد: یکی در آفریقای جنوبی که برای نخستین بار در سال ۲۰۲۱ یک حق اختراع با نام هوش مصنوعی به‌عنوان مخترع صادر کرد (بیدار، ۱۴۰۱)؛ دیگری در استرالیا که دادگاه بدوی ابتدا رأی داد قانون اختراعات این کشور مانعی برای پذیرش مخترع غیر انسان ندارد (بیدار، ۱۴۰۱). هرچند رأی استرالیا در مرحله تجدیدنظر نقض شد و اکنون اکثریت قریب‌به‌اتفاق کشورها بر ضرورت انسان بودن مخترع تأکید دارند (قیصری و همکاران، ۱۴۰۳)؛ اما همین موارد استثنائی، بحث‌های حقوقی دامنه‌داری را برانگیخته است. در حقوق ایران، تاکنون موردی از این دست گزارش نشده و قاعدتاً اگر فردی بکوشد اختراعی را با نام هوش مصنوعی به ثبت برساند، اداره مالکیت صنعتی آن را نخواهد پذیرفت. بر اساس اصول کلی قانون مدنی ایران نیز اهلیت تمتع حقوق صرفاً برای اشخاص (حقیقی و حقوقی) شناخته شده و اموال و حقوق نمی‌توانند بلاصاحب یا متعلق به شیء فاقد شخصیت باشند؛ لذا، هوش مصنوعی که نه انسان است و نه شخص حقوقی، خود نمی‌تواند مالک یا صاحب حق تلقی گردد (قانون مدنی ایران). این اصل کلی در کنار نبود تصریح قانونی، ایجاب می‌کند که حقوق اختراع یا کپی‌رایت خروجی‌های هوش مصنوعی یا به یک شخص واجد اهلیت (مثلاً مخترع/مؤلف انسانی مرتبط) تخصیص یابد یا اساساً ایجاد نشود و اثر در زمره اموال عمومی قرار گیرد.

## تحلیل حقوقی و دیدگاه‌های موجود

با توجه به وضعیت قوانین موجود، حقوقدانان و سیاست‌گذاران سه رویکرد اصلی را در مواجهه با مسئله مالکیت فکری تولیدات هوش مصنوعی مطرح کرده‌اند (Yanisky-Ravid & Vardi, 2017):

۱. توسعه‌دهنده یا مالک سیستم هوش مصنوعی به‌عنوان مالک اثر: بر اساس این دیدگاه، شرکتی که مدل هوش مصنوعی را طراحی و آموزش داده یا به‌طورکلی شخصی که از لحاظ مالی و فنی در ایجاد سیستم نقش

اصلی را داشته است، باید مالک خروجی‌های آن محسوب شود (Gervais, 2019). توجیه این رویکرد آن است که هوش مصنوعی ابزاری در دست سازنده‌اش بوده و محصول نهایی درواقع دستاورد سرمایه‌گذاری و ابتکار اوست. به‌ویژه در مواردی که خروجی سیستم مستقیماً حاصل ساختار و داده‌های آموزش‌دیده آن است، نقش خالقانه کاربر حداقلی به نظر می‌رسد. بااین‌حال، منتقدان این رویکرد استدلال می‌کنند که چنین نگاهی خلاقیت کاربر یا سفارش‌دهنده را نادیده می‌گیرد (Hristov, 2023). چه‌بسا کاربری که با بهره‌گیری خلاقانه از یک هوش مصنوعی اثری بدیع تولید می‌کند، نقشی کمتر از توسعه‌دهنده در پیدایش اثر نداشته باشد. همچنین دادن تمام حقوق به شرکت‌های صاحب فناوری ممکن است به تمرکز قدرت در دست معدودی شرکت بزرگ بیانجامد و انگیزه کاربران فردی را کاهش دهد.

۲. کاربر (طراح پرامپت) به‌عنوان خالق یا مالک اثر: این دیدگاه که طرفداران زیادی یافته، کاربر انسانی را که دستورها، ایده یا داده اولیه را به سیستم می‌دهد شایسته مالکیت اثر می‌داند (Ginsburg & Budiardjo, 2019). استدلال حامیان این نظریه، همان‌طور که در بخش مرور ادبیات آمد، آن است که کار با یک هوش مصنوعی پیشرفته خود نوعی هنر و خلاقیت است؛ انتخاب دقیق کلمات یا پارامترها، اصلاح مکرر ورودی برای رسیدن به نتیجه مطلوب و درنهایت انتخاب بهترین خروجی از میان نتایج به‌دست‌آمده همگی مستلزم ابتکار، ذوق و تلاش انسانی هستند (Yanisky-Ravid & Vardi, 2017). پس اثر نهایی، اگرچه به‌واسطه ماشین تولید شده؛ اما نمود خلاقیت غیرمستقیم انسان است و باید حقوق آن به وی تعلق گیرد. چالش اساسی در این رویکرد آن است که تعیین کنیم چه میزان دخالت و ابتکار از سوی کاربر لازم است تا وی را واقعاً خالق اثر بدانیم (Hristov, 2023). آیا صرف فشردن یک دکمه و تولید تصادفی یک تصویر کافی است یا کاربر باید نقش فعالی در هدایت فرآیند داشته باشد؟ این امر نیازمند تدوین معیارهای حقوقی (و فنی) روشنی است. برخی پیشنهاد کرده‌اند که اگر خروجی هوش مصنوعی قابل پیش‌بینی و تحت کنترل کامل کاربر باشد، آنگاه کاربر خالق محسوب شود؛ اما اگر سیستم به شکل خودگردان و غیرقابل‌پیش‌بینی اثری را پدید آورد (به‌نحوی‌که کاربر تنها ناظر نتیجه است)، در این حالت کاربر را نمی‌توان خالق دانست. ترسیم این مرز در عمل دشوار بوده و احتمال مناقشه در پرونده‌های قضایی را به همراه دارد.

۳. عدم شناسایی مالک خاص (حوزه عمومی): دیدگاه سوم معتقد است زمانی که خالق مستقیم یک اثر انسان نیست، اصولاً هیچ حق انحصاری پدید نمی‌آید و اثر مزبور باید همچون آثار بدون صاحب در دسترس عموم قرار گیرد (Abbott, 2016). طرفداران این نظریه به مبانی سنتی حقوق مؤلف استناد می‌کنند که حمایت انحصاری را ابزاری برای تشویق خلاقیت انسانی می‌دید. در مورد هوش مصنوعی استدلال می‌شود که خود هوش مصنوعی نیازی به تشویق ندارد و توسعه‌دهندگان و کاربران آن نیز انگیزه‌های دیگری (ازجمله منافع تجاری یا علمی) برای استفاده از هوش مصنوعی دارند. پس اگر خروجی‌های هوش مصنوعی به مالکیت عمومی درآیند، دانش و فرهنگ بشری غنی‌تر خواهد شد بدون آن‌که لطمه‌ای به نوآوری وارد شود. انتقاد اصلی به این رویکرد آن است که در عمل ممکن است انگیزه سرمایه‌گذاری در پروژه‌های هوش مصنوعی کاهش یابد (Gervais, 2019). شرکت‌ها و افراد هنگامی روی سامانه‌های خلاق هوش مصنوعی سرمایه‌گذاری سنگین می‌کنند که امید به کسب منافع از نتایج داشته باشند. اگر هیچ‌گونه حمایتی بر نتایج متصور نباشد و فوراً دیگران بتوانند از آن‌ها کپی‌برداری و بهره‌برداری کنند، ممکن است روند نوآوری کند شود. همچنین این دیدگاه افراطی ممکن است عادلانه هم تلقی نشود؛ چراکه نقش غیرمستقیم خالقان و کاربران انسانی را کاملاً بی‌پاداش می‌گذارد.

علاوه بر مباحث مربوط به مالک اثر، جنبه‌های حقوقی دیگری نیز در آثار هوش مصنوعی قابل‌توجه است. یکی از آن‌ها بحث حقوق معنوی[1] پدیدآورنده است که در حقوق ایران و بسیاری کشورها به رسمیت شناخته می‌شود (مثل حق انتساب اثر به نام پدیدآور و حق تمامیت اثر). اعطای این حقوق به هوش مصنوعی بی‌معناست؛ چراکه فاقد شخصیت انسانی است و احساس شرافت و اعتبار ادبی هنری ندارد. از سوی دیگر، اگر بخواهیم کاربر یا توسعه‌دهنده را به‌عنوان پدیدآورنده فرضی معرفی کنیم، آیا مثلاً درست است اثری را که عملاً توسط هوش مصنوعی خلق شده با نام شخص دیگری منتشر کنیم؟ این امر از منظر صداقت علمی و هنری مورد سؤال قرار گرفته است. در عمل البته بسیاری از کاربران خروجی‌های هوش مصنوعی را به نام خود عرضه می‌کنند؛ اما در چارچوب حقوقی، نسبت دادن یک اثر به شخصی که خالق واقعی آن نبوده محل تردیدهای اخلاقی و حقوقی است. برخی حقوقدانان پیشنهاد کرده‌اند که در صورت حمایت از آثار هوش مصنوعی، فقط حقوق مالی اعطا شود و حقوق معنوی سنتی منتفی گردد یا به شکل محدودتری اعمال شود (قیصری و همکاران، ۱۴۰۳؛ Ginsburg & Budiardjo, 2019).

مسئله مهم دیگر، مسئولیت حقوقی در قبال نقض احتمالی حقوق دیگران توسط هوش مصنوعی است. فرض کنیم یک سیستم هوش مصنوعی در فرآیند یادگیری خود از آثار دارای کپی‌رایت بدون اجازه استفاده کرده و سپس خروجی‌ای تولید کند که مشابه اثر اصلی است. چنین موردی در حال حاضر موضوع دعاوی حقوقی مهمی قرار گرفته است؛ برای نمونه شرکت Getty Images و گروهی از هنرمندان از توسعه‌دهندگان مدل‌های مولد تصویر (مانند Stable Diffusion) شکایت کرده‌اند که چرا آثار دارای کپی‌رایت آن‌ها را برای آموزش الگوریتم استفاده نموده و سپس تصاویری شبیه به آن سبک‌ها تولید می‌کنند (Hristov, 2023). این دعاوی بحث را فراتر از مالکیت خروجی‌ها به مشروعیت داده‌های ورودی و روش یادگیری هوش مصنوعی کشانده است. هرچند موضوع یادشده خود مجال مفصل دیگری می‌طلبد؛ اما ارتباط آن با بحث حاضر در این است که خروجی‌های هوش مصنوعی ممکن است به‌طور ناخودآگاه ناقض حقوق دیگر صاحبان آثار باشند. در چنین حالتی چه کسی مسئول شناخته می‌شود؟ کاربری که فرمان تولید را داده، یا شرکتی که الگوریتم را ساخته، یا هیچ‌کدام (با این استدلال که عمل مستقیماً از سوی ماشین سر زده است)؟ پاسخ به این سؤالات هنوز در مراحل ابتدایی بحث است ولی احتمالاً نظام‌های حقوقی همان رویه کلی را پیش خواهند گرفت که در سایر موارد مسئولیت محصولات و ابزارها دارند: یعنی مسئولیت غیرمستقیم یا ناشی از تقصیر متوجه انسان‌هایی خواهد بود که کنترل یا بهره‌برداری از هوش مصنوعی را بر عهده داشته‌اند (مانند تولیدکننده یا استفاده‌کننده از محصول معیوب). در هر صورت، چالش‌های اجرایی مرتبط با احراز مصادیق تخلف و تعیین مسئولیت در حوزه آثار هوش مصنوعی بسیار پیچیده‌تر از حالت سنتی بوده و نیازمند توسعه قواعد روشنی در آینده است (Yanisky-Ravid & Vardi, 2017).

## چالش‌های اجرایی و حقوقی

بررسی‌های بالا نشان می‌دهد که خلأ قانونی و ابهام مفهومی در زمینه مالکیت فکری آثار هوش مصنوعی می‌تواند به چالش‌های عملی متعددی منجر شود. در این بخش به برخی از مهم‌ترین این چالش‌ها در دو بُعد اجرایی و حقوقی اشاره می‌کنیم:

- عدم قطعیت حقوقی و افزایش منازعات: نبود قوانین شفاف، خود می‌تواند انگیزه طرح دعاوی و منازعات را افزایش دهد. برای مثال، اگر دو فرد مختلف مدعی مالکیت یک اثر تولیدشده توسط هوش مصنوعی شوند (یکی به‌عنوان سازنده الگوریتم و دیگری به‌عنوان کاربر خالق اثر)، دادگاه‌ها باید بر اساس قیاس از قوانین

1. moral rights

موجود یا دکترین‌های حقوقی عمومی تصمیم بگیرند که ممکن است در پرونده‌های مشابه نتیجه متفاوتی بدهند. چنین وضعیتی موجب سردرگمی ذی‌نفعان و کاهش قابلیت پیش‌بینی حقوقی می‌شود که خود مانعی بر سر راه نوآوری است (Gervais, 2019). همچنین، اگر رویه قضایی روشنی شکل نگیرد، ممکن است برخی افراد یا شرکت‌ها از خلأ قانونی سوءاستفاده کرده و حقوق دیگران را تضییع کنند (مثلاً با کپی و بهره‌برداری آزادانه از آثار هوش مصنوعی تولید دیگران به این بهانه که حمایت‌شده نیستند). درنتیجه، هرگونه تأخیر در اصلاح مقررات می‌تواند تبعات عملی ناگواری داشته باشد.

- مشکلات اداری در ثبت و حفاظت از حقوق: نظام‌های ثبتی فعلی (مانند ادارات ثبت اختراع یا سیستم ثبت و شماره‌دهی آثار در وزارت ارشاد برای کتاب‌ها و نرم‌افزارها) بر محور اطلاعات هویتی انسان‌ها طراحی شده‌اند. اگر امروز فردی برای ثبت یک اثر هنری تولیدشده توسط هوش مصنوعی مراجعه کند و در فرم تقاضا در قسمت پدیدآورنده بنویسد «یک برنامه هوش مصنوعی»، احتمالاً درخواست وی رد خواهد شد (بیدار، ۱۴۰۱). همچنین در ثبت اختراع، ذکر نام مخترع انسانی الزامی است (قیصری و همکاران، ۱۴۰۳). این امر باعث می‌شود برخی افراد به‌ناچار اطلاعات خلاف واقع ارائه دهند (مثلاً نام خود یا همکارشان را به‌جای هوش مصنوعی به‌عنوان مخترع ذکر کنند) تا اختراعشان قابل ثبت شود. چنین راهکارهایی هرچند حقوقی را ظاهراً اعطا می‌کند ولی از نظر اصول اخلاقی و شفافیت حقوقی مطلوب نیست و می‌تواند در آینده اعتبار حقوق ثبت‌شده را زیر سؤال ببرد (برای مثال، اگر افشا شود که مخترع واقعی انسان نبوده است)؛ بنابراین، سازوکارهای اداری نیاز به تطبیق و به‌روزرسانی دارند تا بتوانند شرایط خاص آثار هوش مصنوعی را در نظر بگیرند، خواه از طریق فرم‌های جدید یا دستورالعمل‌های اجرایی صریح.
- ارزیابی معیارهای اصالت و خلاقیت: در حقوق کپی‌رایت، اصالت[1] شرط حمایت اثر است. ارزیابی اصالت معمولاً به این معنی است که اثر برآمده از تلاش و خلاقیت مستقل پدیدآورنده باشد و نه کپی از دیگری. در مورد آثار هوش مصنوعی یک سؤال پیچیده آن است که اصالت را به چه کسی منتسب کنیم؟ اگر انسانِ کاربر ایده کلی یا سبک را تعیین کرده و هوش مصنوعی جزئیات را پردازش کرده باشد، آیا اثر حاصل اصیل است؟ از یک طرف بله، چون خروجی منحصربه‌فرد و نو است؛ اما از طرف دیگر، چون مستقیماً زاییده ذهن انسان نیست، ممکن است اصیل تلقی نشود (Ginsburg & Budiardjo, 2019). همچنین، با پیشرفت هوش مصنوعی، تمییز دادن آثار هوش مصنوعی از آثار انسان‌ساخته دشوارتر می‌شود. این مسئله برای ممیزان ادارات ثبت و حتی دادگاه‌ها چالش‌برانگیز خواهد بود که خط تمایز میان اثر اصیل واجد حمایت و صرف ترکیب مکانیکی داده‌ها را تشخیص دهند.
- حجم انبوه آثار تولیدی و امکان سوءاستفاده: یکی از ویژگی‌های هوش مصنوعی مولد آن است که می‌تواند در زمان کوتاهی حجم عظیمی از خروجی‌ها تولید کند. تصور کنید یک نفر با استفاده از هوش مصنوعی ظرف یک روز هزاران تصویر یا متن مختلف ایجاد کند. اگر این‌ها قابل حمایت باشند، آیا وی می‌تواند ادعای حق مؤلف بر همه آن‌ها داشته باشد؟ این امر می‌تواند بار سنگینی بر سیستم ثبت و اجرای حقوق مالکیت فکری تحمیل کند و مفهوم منحصر بودن اثر را که به‌طور سنتی با تلاش و زحمت پدیدآورنده گره خورده بود، مخدوش نماید. همچنین نگرانی از سوی هنرمندان و نویسندگان وجود دارد که فضای رقابت غیر یکسان شود؛ به این معنی که افرادی با کمک هوش مصنوعی و بدون صرف زمان خلاقانه زیاد، انبوهی اثر تولید کنند

1. originality

و حقوق آن را بگیرند، درحالی‌که خالقان انسانی برای هر اثر باید زمان و نبوغ صرف کنند. این نیز از جنبه سیاست‌گذاری فرهنگی و حمایت از خلاقیت انسانی دغدغه‌آفرین است (Hristov, 2023).

- بعد فرامرزی و تعارض قوانین: درنهایت باید توجه داشت که مسئله آثار هوش مصنوعی یک پدیده جهانی است و خروجی‌های یک سیستم می‌تواند به‌سرعت در سراسر دنیا منتشر شود. اختلاف رویکرد کشورها (مثلاً اگر کشوری اثری را فاقد حمایت بداند و کشور دیگر آن را مشمول کپی‌رایت) می‌تواند به آشفتگی در رعایت حقوق میان کشورها بینجامد. برای نمونه، ممکن است اثری که در ایران تولید شده و اینجا حمایت نشده، توسط شخصی در اروپا ثبت و ادعای حقوق شود یا بالعکس. چنین وضعیتی ضرورت هماهنگی‌های بین‌المللی و گفت‌وگوی مستمر میان قانون‌گذاران را یادآور می‌شود تا حتی‌الامکان استانداردهای مشترکی برای تعیین تکلیف این آثار پیدا شود (WIPO, 2020).

## نتیجه‌گیری و پیشنهادها

مالکیت فکری در تولیدات خودکار هوش مصنوعی یکی از چالش‌برانگیزترین موضوعات حقوقی عصر حاضر است که ابعاد حقوقی، اقتصادی و فلسفی گسترده‌ای دارد (Abbott, 2016). از یک سو، نادیده گرفتن و حمایت نکردن از این قبیل آثار می‌تواند به هرج‌ومرج و کاهش انگیزه نوآوران منجر شود؛ چراکه افراد و شرکت‌ها در صورت نبود حمایت حقوقی ممکن است تمایلی به انتشار یا سرمایه‌گذاری در پروژه‌های خلاق مبتنی بر هوش مصنوعی نداشته باشند (Gervais, 2019). از سوی دیگر، اعطای حمایت کامل و برابر با آثار کاملاً انسانی نیز بدون مخاطره نیست؛ زیرا بیم آن می‌رود که ارزش و جایگاه خلاقیت انسانی تنزل یابد و همچنین سیل آثار آسان‌تولیدشده توسط ماشین، نظام مالکیت فکری را با حجم غیرقابل‌کنترلی مواجه سازد (Hristov, 2023)؛ بنابراین، راهکار مطلوب، اتخاذ رویکردی است که نه مانع پیشرفت فناوری شود و نه دستاوردهای آن را به حال خود رها کند بلکه به شکلی نظم‌دهنده و آینده‌نگر به تنظیم حقوق و تکالیف بپردازد (Ginsburg & Budiardjo, 2019).

برای نظام حقوقی ایران که قوانین فعلی‌اش پاسخگوی صریح این معضل نیستند (فتاحی، ۱۴۰۲)، اقدامات زیر پیشنهاد می‌گردد:

۱. بازنگری قانون حقوق مؤلف و قوانین مرتبط: ضروری است قانون‌گذار با تشکیل کارگروه‌های تخصصی، قانون ۱۳۴۸ و سایر مقررات را به‌روز کرده و تکلیف آثار بدون خالق انسانی را مشخص نماید. یکی از گزینه‌ها الهام گرفتن از تجربه بریتانیا است، به این نحو که قانون تصریح کند در صورت فقدان خالق انسانی، شخصی که نقش تعیین‌کننده در خلق اثر ایفا کرده (اعم از برنامه‌نویس یا کاربر یا سرمایه‌گذار اصلی) به‌عنوان پدیدآورنده انگاشته شود. البته تعیین مصداق این شخص در هر مورد می‌تواند دشوار باشد و شاید نیاز به آیین‌نامه اجرایی یا دستورالعمل تکمیلی داشته باشد (قیصری و همکاران، ۱۴۰۳).

۲. ایجاد حق ویژه یا شبه‌کپی‌رایت برای آثار هوش مصنوعی: همان‌گونه که برخی متخصصان پیشنهاد کرده‌اند (Yanisky-Ravid & Vardi, 2017)، می‌توان یک دسته جدید در نظام مالکیت فکری تعریف کرد که مختص آثار تولیدشده توسط سامانه‌های هوشمند باشد. این حق می‌تواند از حیث مدت حمایت و دامنه حقوق محدودتر از کپی‌رایت معمولی باشد تا هم انگیزه لازم برای نوآوری باقی بماند و هم انحصار کوتاه‌مدت مانع بهره‌برداری عمومی طولانی‌مدت نشود. به‌عنوان‌مثال، شاید حمایت ۵ یا ۱۰ ساله غیرقابل‌تمدید برای این آثار کافی باشد. همچنین، می‌توان برخی حقوق معنوی نظیر حق انتساب را تعدیل کرد (مثلاً الزام به ذکر تولید توسط هوش مصنوعی در کنار نام صاحب حق).

۳. تدوین رویه در حوزه اختراعات هوش مصنوعی: پیشنهاد می‌شود اداره ثبت اختراعات ایران، با رصد تحولات جهانی، دستورالعملی صادر کند که مشخص نماید اگر اختراعی با دخالت هوش مصنوعی ایجاد شده، نحوه اظهار مخترع چگونه باشد. شاید بهترین رویکرد کماکان اصرار بر ذکر نام یک یا چند مخترع انسانی است (به‌عنوان افرادی که ایده اولیه یا تنظیم پروژه را انجام داده‌اند) تا اختراع ثبت‌شده از نظر قانونی متزلزل نشود. درعین‌حال لازم است در فرم‌های ثبت اختراع بخشی اضافه شود که میزان استفاده از هوش مصنوعی در ابداع را توضیح دهد. این شفافیت کمک می‌کند در آینده اگر قانون تغییر کرد یا اختلافی بروز نمود، سابقه روشنی از نقش هوش مصنوعی در پرونده اختراع موجود باشد (بیدار، ۱۴۰۱).
۴. آموزش و توسعه ظرفیت کارشناسی: سیستم قضایی و اداری برای مواجهه با چنین پرونده‌های نوینی نیاز به ارتقای دانش فنی و حقوقی دارد. آموزش قضات، کارشناسان دادگستری و ممیزان ادارات مالکیت فکری در زمینه مبانی فناوری هوش مصنوعی و روش کار آن‌ها ضروری است تا بتوانند تفکیک میان سهم انسان و ماشین در خلق آثار را بهتر درک کنند. همچنین دعوت از صاحب‌نظران فناوری در تدوین قوانین و مقررات جدید به حصول راهکارهای عملیاتی‌تر منجر خواهد شد (WIPO, 2020).
۵. همکاری‌های بین‌المللی: مسئله هوش مصنوعی مرز نمی‌شناسد و به همین جهت لازم است ایران در مجامع بین‌المللی (مانند WIPO) به شکل فعال حضور داشته باشد تا در روند شکل‌گیری قواعد و رویه‌های جهانی نقش ایفا کند. تبادل تجربیات با دیگر کشورها به‌ویژه کشورهای دارای نظام حقوقی مشابه (مثلاً کشورهای دارای حقوق نوشته و الهام‌گرفته از حقوق فرانسه) می‌تواند مفید باشد. شاید در آینده یک معاهده بین‌المللی یا حداقل توصیه‌نامه‌هایی درباره حقوق آثار هوش مصنوعی تدوین شود که بهتر است ایران جزو پیشگامان آن باشد تا منافع ذی‌نفعان داخلی نیز لحاظ گردد (WIPO, 2020).

در پایان، تأکید می‌شود که عصر هوش مصنوعی علاوه بر چالش‌ها، فرصت‌هایی نیز برای نظام حقوقی به همراه دارد. همان‌گونه که فناوری‌های پیشین (چاپ، عکاسی، رایانه و اینترنت) موجب تحول در نگرش‌های حقوقی شدند و نهایتاً قوانین با آن‌ها تطبیق یافت، هوش مصنوعی نیز حقوق را به تکاپو برای نوآوری و پویایی واداشته است. تنظیم‌گری مناسب در این حوزه می‌تواند ضمن حمایت از رشد فناوری و خلاقیت‌های نو، ارزش‌های بنیادین حقوق مالکیت فکری یعنی تکریم خلاقیت انسان و تشویق نوآوری را در سطحی جدید استمرار بخشد (Abbott, 2016).

## منابع